# Global parameter optimization of Mather type plasma focus in the framework of the Gratton-Vargas two-dimensional snowplow model

S K H Auluck
Physics Group, Bhabha Atomic Research Center, Mumbai 400085 INDIA

skauluck@barc.gov.in; skhauluck@gmail.com

**Abstract.** Dense Plasma Focus (DPF) is known to produce highly energetic ions, electrons and plasma environment which can be used for breeding of short-lived isotopes, plasma nanotechnology and other material processing applications. Commercial utilization of DPF in such areas would need a design tool which can be deployed in an automatic search for the best possible device configuration for a given application. The recently revisited [S K H Auluck, Physics of Plasmas 20, 112501 (2013)] Gratton-Vargas (GV) two-dimensional analytical snowplow model of plasma focus provides a numerical formula for dynamic inductance of a Mather type plasma focus fitted to thousands of automated computations, which enables construction of such design tool. This inductance formula is utilized in the present work to explore global optimization, based on first-principles optimality criteria, in a 4-dimensional parameter-subspace of the zero-resistance GV model. The optimization process is shown to reproduce the empirically observed constancy of the drive parameter over 8 decades in capacitor bank energy. The optimized geometry of plasma focus normalized to anode radius is shown to be independent of voltage, while the optimized anode radius is shown to be related to capacitor bank inductance.

## 1. Introduction:

Optimized design of dense plasma focus (DPF) devices has been a long standing goal since the early days of DPF research [1]. Many empirical criteria and experimental optimization procedures have been suggested over the years [2-7] for choosing the device parameters for a given capacitor bank. Simplified numerical models [8-15] have been deployed in the quest for a well-optimized device. One of the observations [16-22] of this research is that the drive parameter, $I/a\sqrt{p}$, where I is the peak current, a is the anode radius and p is the deuterium gas pressure, has nearly constant value for devices experimentally optimized for neutron production. This implies [16-19] that many parameters such as axial and radial velocities, ion temperature, plasma energy density, Alfven velocity, magnetic energy per unit mass have nearly identical values across a wide spectrum of neutron-optimized devices spanning 8 decades of capacitor bank energy. Many possible physical reasons have been proposed [1,19,20] for this circumstance.

Designing a DPF device for a given capacitor bank, characterized by the four parameters - capacitance, inductance, resistance and voltage, involves choosing values for the following 6 parameters –anode radius, anode length, insulator radius, insulator length, cathode radius and gas pressure (or density). The cathode length is generally taken to be equal to the anode length; a somewhat different value is sometimes adopted for facilitating diagnostic access but usually that does not lead to a noticeable difference in the device properties. The empirically observed near-constancy of drive parameter in neutron-optimized devices is suggested as a design criterion for determining the anode radius [21,22] using the additional empirical observation that most DPF devices work with a few millibar pressure of deuterium; the anode length and working pressure are then chosen so that the plasma arrives at the axis at the peak of current [22]. Recent realization of DPF [23] which operates at unusually high deuterium pressures of tens of millibar and which also reports a better-than-global-scaling neutron yield [23] raises the possibility that conventionally-designed neutron-optimized DPF devices may not represent a globally optimized ("best possible") DPF device.

Resurgence of interest over the last 10 years [24] in commercially significant applications of DPF [25,26,27] indicates that the time has arrived for a deeper examination of the question of global optimization of DPF in the 6-dimensional parameter space, which is based on a transparent model of DPF operation devoid of unstated assumptions. Of particular importance are applications for plasma nanotechnology processes [25], where DPF acts as a provider of a unique plasma environment rather than of fusion neutrons. Breeding short-lived isotopes [28] for medical diagnostics is also of considerable commercial interest. These two applications use fundamentally different properties of the DPF: the former uses the intense power delivered by soft-x-rays (for lithography) or plasma and ions with few tens of eV temperature /kinetic energy (for coatings or surface treatment); the latter uses confined ions with hundreds of keV energy interacting with relatively dense target plasma. Industrial-scale investment in development of DPF as a technology platform for commercial utilization of these phenomena would demand that the adopted design should be globally optimized ("best possible") for the intended application in order to avoid the risk of premature technical obsolescence, using an experimentally validated design tool and well-defined, first-principles optimization criteria, not excessively dependent on but compatible with empirical thumb rules, in an automated unbiased parameter search.

The practical logic of industrial-scale investment has significant implications for scientific aspects of global optimization efforts. A purely scientific view of optimization would involve maximization of appropriately defined quantitative performance criteria subject to known technical constraints. A commercial view of optimization would include the possibility of overcoming some of the technical constraints through innovation, (such as new kinds of current generators, new ways of forming the initial plasma, new device geometries), which would have the effect of protecting the technical leadership of the investor through intellectual property rights. It would also involve strategic trade-offs, sacrificing a technically better option in favor of one that affords long term business advantages or which makes better commercial sense in the short term. This implies that techno-commercial optimality criteria themselves are undefined *a priori*; they are to be 'discovered' iteratively as part of the optimization project. Therefore, the optimization effort needs to be based on a non-judgmental *tabulation* of the behavior of a variety of optimality parameters in practically important regions of the parameter space followed by a process of discovery of the optimality conditions and of the optimal configuration itself.

Recent re-appraisal [29] of the Gratton-Vargas (GV) two-dimensional analytical snowplow model of plasma focus evolution [30] has revealed opportunities for global parametric optimization of the Mather type plasma focus based on a numerical formula for its dynamic inductance, determined by fitting inductance data calculated from thousands of automated computations. Current profile data from contemporary DPF facilities can be fitted very well with the proposed modification [29] of the GV model to include circuit resistance, when gas fill pressure, static inductance and circuit resistance are treated as fitting parameters. This formulation enables calculation of certain quantities related to the current profile in a very short time, enabling automated tabulation of optimal properties of DPF configurations in the 6-dimensional parameter space. The present work seeks to initiate exploration of these opportunities.

The next section recapitulates relevant results from the revised Resistive GV model [29]. Section 3 introduces the concept of similarity classes of the GV model and looks at their properties. Some issues involved with global parametric optimization are described in section 4. Examples of automated parameter space survey using an optimization algorithm are described in section 5. Section 6 presents a summary of the main results and conclusions.

## 2. Salient features of the revised GV model:

This section recapitulates nomenclature and salient features of the revised resistive GV model [29], in order to provide a condensed, self-contained background for the present discussion. The GV model is based on the snowplow hypothesis, which equates the magnetic pressure acting behind the plasma current sheath (PCS) with the 'wind pressure' experienced by the PCS driven into stationary neutral gas. This results in a partial differential equation (called GV equation [29]) for the propagation of the (azimuthally symmetric) PCS in two-dimensional (r,z) space as a function of time. The GV equation admits scaling to a dimensionless form, where coordinates (r,z) and linear dimensions are expressed in units of the anode radius 'a': $\tilde{r} \equiv r/a$ and $\tilde{z} \equiv z/a$ and time is replaced as an independent variable by the dimensionless variable

$$\tau(t) = \frac{1}{Q_m} \int_0^t I(t') dt' \qquad 1$$

where $Q_m \equiv \pi a^2 \sqrt{2\mu_0 \rho_0}/\mu_0$ is a quantity having dimensions of charge ('mechanical equivalent of charge'), $I(t)$ is capacitor bank discharge current as a function of time t and $\rho_0$ is the mass density of the fill gas, all quantities in SI units. The GV equation can be solved analytically using the method of characteristics and the shape and location of the PCS, $\tilde{z} = \overline{f}(\tilde{r}, \tau)$, can be determined [29] as a function of the independent variable $\tau$ for a Mather-type DPF, assumed to have a straight solid cylinder of radius 1 and length $\tilde{z}_A$ as anode, a straight cylinder of outer radius $\tilde{r}_I$ and length $\tilde{z}_I$ as insulator and a straight cylinder of inner radius $\tilde{r}_C$ and height $\tilde{z}_A$ as cathode, in terms of two characteristic values of $\tau$: $\tau_{LIFTOFF} = \tilde{r}_C^2 - \tilde{r}_I^2$, $\tau_R = 2(\tilde{z}_A - \tilde{z}_I)$. The PCS reaches the device axis at $\tau = \tau_R + 1$. However, at an empirically determined[17] 'pinch radius' $\tilde{r}_p \simeq 0.12$, the assumption of snowplow model breaks down, and therefore, the GV model is considered to be valid only up to $\tau^* \simeq \tau_R + 0.98$. The GV model is expected to be progressively less accurate as this stage is approached because of the increasingly important role played by gas dynamic phenomena neglected in the model.

From this description of the PCS, the dynamic plasma inductance can be calculated from the magnetic flux enclosed between the current sheath and the electrodes as a function of $\tau$

$$L_P(\tau) = \frac{\mu_0 a}{2\pi} \iint d\tilde{z} d\tilde{r} \frac{1}{\tilde{r}} \equiv \frac{\mu_0 a}{2\pi} \mathcal{L}(\tau) \qquad 2$$

The current $I(t)$ obeys the following circuit equation for a capacitor bank of capacitance $C_0$, internal inductance $L_0$ and internal resistance $R_0$ charged to voltage $V_0$:

$$\frac{d}{dt}(LI) = V_0 - \frac{1}{C_0}\int_0^t I(t')dt' - IR_0 \qquad 3$$

Introducing dimensionless quantities:

$\varepsilon \equiv Q_m / C_0 V_0$, $\kappa \equiv \mu_0 a / 2\pi L_0$, $\gamma \equiv R_0 \sqrt{C_0/L_0}$, $\tilde{I}(\tau) \equiv I(\tau(t))/I_0$,

$\Phi \equiv (LI)/L_0 I_0 = (1 + \kappa \mathcal{L}(\tau))\tilde{I}$ where $I_0 \equiv V_0 \sqrt{C_0/L_0}$, this takes the form

$$\Phi \frac{d\Phi}{d\tau} = \varepsilon(1 + \kappa \mathcal{L}(\tau))(1 - \varepsilon\tau) - \varepsilon\gamma\Phi \qquad 4$$

For the case of zero capacitor bank resistance ($\gamma = 0$), 4 gives

$$\Phi_0^2 = 2\varepsilon \int_0^\tau d\tau' \left(1 + \mathcal{L}(\tau')\kappa\right)(1 - \varepsilon\tau')$$

$$= 2\varepsilon\tau - \varepsilon^2\tau^2 + 2\varepsilon\kappa m_0(\tau) - 2\varepsilon\kappa\varepsilon m_1(\tau) \quad \quad 5$$

$$m_0(\tau) \equiv \int_0^\tau d\tau' \mathcal{L}(\tau'); m_1(\tau) \equiv \int_0^\tau d\tau' \tau' \mathcal{L}(\tau')$$

The case of non-zero circuit resistance is dealt with using the method of successive approximations using the smallness of the parameter $\varepsilon\gamma$. The flux function $\Phi(\tau)$ is treated as the limit of a sequence of functions $\Phi_n(\tau)$, $n = 0, 1, 2\cdots$ obeying the equation

$$\Phi_{n+1}\frac{d\Phi_{n+1}}{d\tau} = \varepsilon\left(1 + \mathcal{L}(\tau)\kappa\right)(1 - \varepsilon\tau) - \varepsilon\gamma\Phi_n$$

$$\Rightarrow \Phi_{n+1}^2(\tau) = \Phi_0^2(\tau) - 2\varepsilon\gamma\int_0^\tau d\tau \Phi_n \quad \quad 6$$

The real time t corresponding to the independent variable $\tau$ is determined in terms of the short-circuit quarter-cycle time $T_{1/4} \equiv \pi/2 \cdot \sqrt{C_0 L_0}$:

$$\tilde{t} \equiv t/T_{1/4} = (2\varepsilon/\pi) \cdot \int_0^\tau d\tau'/\tilde{I}(\tau') = (2\varepsilon/\pi) \cdot \int_0^\tau d\tau'\left(1 + \kappa\mathcal{L}(\tau)\right)/\Phi(\tau') \quad \quad 7$$

The partitioning of stored energy $W_0 \equiv \tfrac{1}{2}C_0 V_0^2$ between magnetic energy $W_M$, electromagnetic work $W_W$, dissipation $W_R$ in circuit resistance and energy $W_C$ remaining in the capacitor is described by the relations

$$\eta_M \equiv W_M/W_0 \equiv \frac{\tfrac{1}{2}L(t)I^2(t)}{\tfrac{1}{2}C_0 V_0^2} = \Phi(\tau)^2/\left(1 + \kappa\mathcal{L}(\tau)\right) \quad \quad 8$$

$$\eta_W \equiv W_W/W_0 \equiv \frac{1}{\tfrac{1}{2}C_0 V_0^2}\int \tfrac{1}{2}I^2 dL = \varepsilon\tau(2 - \varepsilon\tau) - \eta_m - 2\varepsilon\gamma\int d\tau \tilde{I} \quad \quad 9$$

$$\eta_R \equiv W_R/W_0 \equiv \frac{R\int dt I^2(t)}{\tfrac{1}{2}C_0 V_0^2} = 2\varepsilon\gamma\int d\tau \tilde{I} \quad \quad 10$$

$$\eta_C \equiv W_C/W_0 = \frac{1}{2C_0}\left(C_0 V_0 - \int_0^t I(t')dt'\right)^2 = (1-\varepsilon\tau)^2 \qquad 11$$

A numerical formula for the dynamic plasma inductance profile $\mathfrak{L}(\tau)$, calculated from the PCS shape using 2, has been fitted [29] to $\mathfrak{L}(\tau)$ data from several thousand automated computations over the range $1.01 \leq \tilde{r}_I \leq 1.04$, $0.5 \leq \tilde{z}_I \leq 2$, $\tilde{r}_I + 0.2 \leq \tilde{r}_C \leq 2.0$,

$$\text{Max}\left[2, \tilde{r}_C, 0.5\tilde{r}_C^2 - 1 + \tilde{r}_I\right] \leq \tilde{z}_A \leq 10 + \tilde{r}_I:$$

$$\begin{aligned}
\mathfrak{L}(\tau) &= \tilde{z}_I \text{Log}\left(\sqrt{\tilde{r}_I^2 + \tau}\right) + k_1 \text{Log}(\tilde{r}_C)\tau^{1.5} & 0 < \tau \leq \tau_{\text{LIFTOFF}} \\
&= \mathfrak{L}(\tau_{\text{LIFTOFF}}) + \frac{1}{2}(\tau - \tau_{\text{LIFTOFF}})\text{Log}(\tilde{r}_C) + k_2 \text{Log}(\tilde{r}_C) & \tau_{\text{LIFTOFF}} \leq \tau \leq \tau_R \qquad 12 \\
&= \mathfrak{L}(\tau_R) - k_3 \text{Log}(\tau_R + 1 - \tau) & \tau_R \leq \tau \leq \tau_R + 0.98
\end{aligned}$$

The three parameters $k_1$, $k_2$, $k_3$ are found to be independent of the scaled anode length $\tilde{z}_A$, scaled insulator length $\tilde{z}_I$ and scaled insulator radius $\tilde{r}_I$. They depend on the scaled cathode radius $\tilde{r}_C$ as

$$k_1 = \frac{\lambda_0}{\tilde{r}_c + \lambda_1}; k_2 = \lambda_2 + \lambda_3 \tilde{r}_c + \lambda_4 \tilde{r}_c^2; k_3 = \lambda_5 + \lambda_6 \tilde{r}_c + \lambda_7 \tilde{r}_c^2 \qquad 13$$

$\lambda_0$=0.276304;  $\lambda_1$=0.68924;  $\lambda_2$=0.08367;  $\lambda_3$=0.105717;  $\lambda_4$=0.02786;  $\lambda_5$=0.05657;  $\lambda_6$=0.263374; $\lambda_7$=0.04005.

The formula for the volume swept by the PCS up to $\tau_R$ [30] in the notation of this paper is given by

$$\upsilon = \pi a^3 \left(\tilde{r}_c^2 - 1\right)\left(\tilde{z}_A - \tfrac{3}{7}h\right); h = \tfrac{1}{2}\tilde{r}_c \sqrt{\tilde{r}_c^2 - 1} - \text{Log}\left(\tilde{r}_c + \sqrt{\tilde{r}_c^2 - 1}\right) \qquad 14$$

It should be noted that the region over which formula 12 has been fitted defines the region for parametric search in this paper; however, it is in principle possible to obtain similar formulas applicable in other regions or indeed for other geometries such as the Filippov geometry [31], hypocycloidal pinch [32] and many still-to-be-discovered concepts.

The GV model has been experimentally verified [33] and used by M. Milanese and co-workers [34-37] and H. Bruzzone and co-workers [38] to interpret their experimental results and generate insight into the dynamics of energy transfer during the rundown phase in the Dense Plasma Focus. A good (manual) fit of the resistive GV (RGV) model with an experimental current waveform has been demonstrated [29] earlier. A refinement of the fit can be obtained by minimizing the expression for the mean deviation

$$\beta = \frac{1}{N}\sqrt{\sum_{i=1}^{N}\left(\tilde{I}_{GV}\left(\tilde{t}_i\right) - \tilde{I}_{exp}\left(\tilde{t}_i\right)\right)^2} \qquad 15$$

defined in terms of experimental digitized current waveform $I_{exp}(t_i)$ normalized to $I_0 \equiv V_0\sqrt{C_0/L_0}$ and $T_{1/4} \equiv \pi/2 \cdot \sqrt{C_0 L_0}$ where both scale factors contain $L_0$, one of the parameters being fitted. This fit extends only over those experimental points (N in number) which fall within the limit of validity $\tau^*$ of the GV model. Fig. 1 shows the best fit corresponding to fig 3 of Ref [29] which minimizes $\beta$ as a function of 3 fit parameters, pressure, inductance, resistance, using automated numerical calculations.

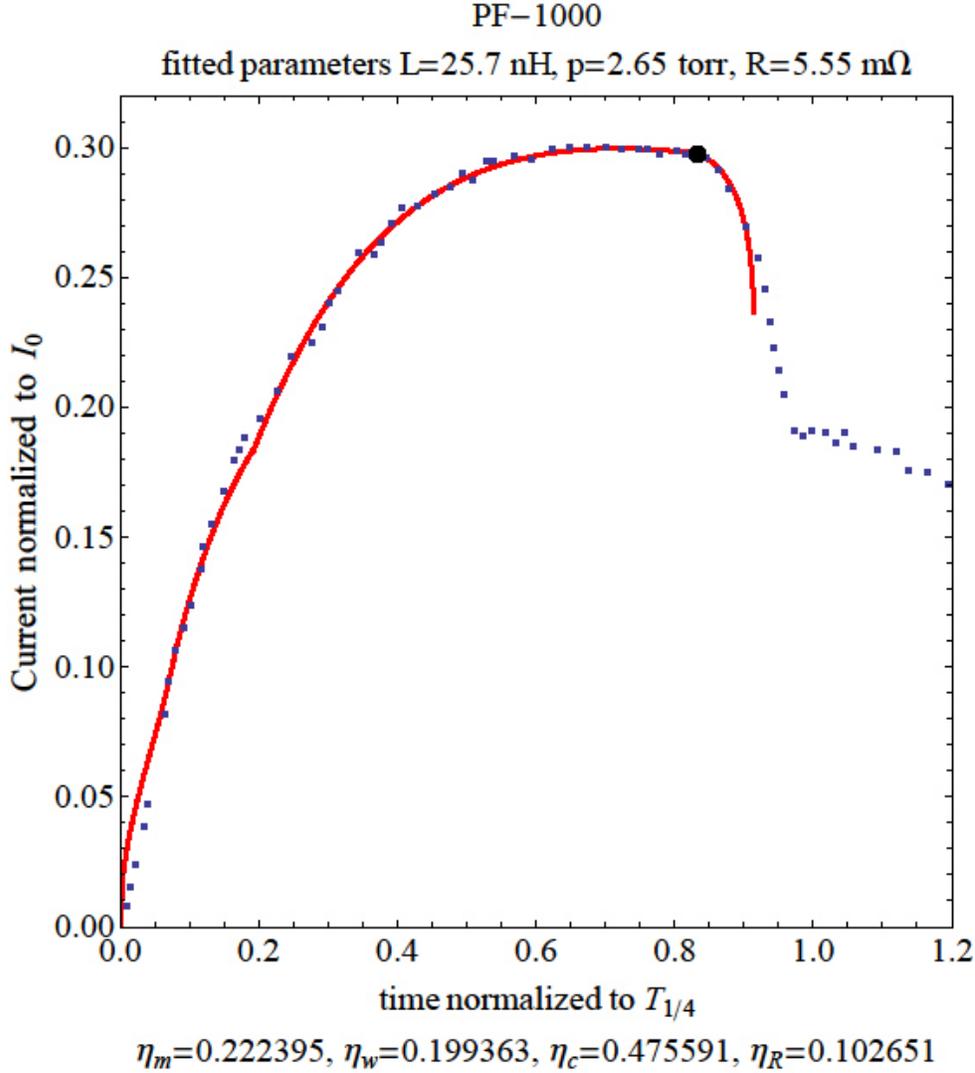

Fig.1: Comparison of RGV model "best fit" (continuous line) with manually digitized experimental data (points) from PF-1000 (see Ref 29, Fig 3 for details). The mean deviation parameter has value $\beta = 0.00096912$. At the limit of validity of the GV model, $\tau^* = \tau_R + 0.98$ which corresponds to $\tilde{t}^* = 0.9156$, the scaled current from the RGV model is 0.2364 while the experimental value is 0.2606. The agreement is within 10% which is the expected combined error of experimental measurements and manual digitization from a printed figure. The model faithfully reproduces the experimental current till the end of the rundown and into the radial phase up to $\tau = \tau_R + 0.822535$, which corresponds to $\tilde{t}^* = 0.900658$, when the sheath reaches a scaled radius ~0.42.

The good quality of fit supports the idea that the resistive GV model proposed earlier [29] is an adequate representation of the gross energy transfer from capacitor bank to the plasma using non-dimensional parameters in an analytical format. It can thus be used as a tool for examination of questions related to the optimization of energy transfer in device performance.

### 3. Similarity classes in the GV model:

A significant feature of the GV model is that the ten parameters describing a DPF facility – capacitance $C_0$, inductance $L_0$, resistance $R_0$, voltage $V_0$, gas pressure p (or density $\rho_0$), anode radius a, anode length, insulator radius, insulator length and cathode radius – are mapped on to 7 independent dimensionless parameters of the RGV model: $\varepsilon \equiv Q_m/C_0V_0$, $\kappa \equiv \mu_0 a/2\pi L_0$, $\gamma \equiv R_0\sqrt{C_0/L_0}$, $\tilde{z}_A, \tilde{z}_I, \tilde{r}_I, \tilde{r}_C$. Each point in the 7-dimensional RGV parameter space then has three degrees of freedom in the space of all possible DPF facilities, which may be chosen to be capacitance $C_0$, internal inductance $L_0$ and charging voltage $V_0$. A similarity class in the GV model is defined as a set of all DPF facilities which have identical values for all dimensionless parameters of the GV model.

All DPF facilities in the GV similarity class (GVSC) are characterized by identical dynamic plasma inductance profile as well as scaled current profile parametrically defined as $\{\tilde{t}(\tau), \tilde{I}(\tau)\}, \tau = 0 \ldots \tau_R + 0.98$. However, they differ in terms of their quantitative performance as well as cost.

One basic distinction between different members of the GVSC consists of the capacitor bank impedance $Z_0 \equiv \sqrt{L_0/C_0}$. Although $\gamma \equiv R_0\sqrt{C_0/L_0}$ has been assumed to be identical for all members of the GVSC, later discussion of Section 4 reveals another characteristic impedance related to physics external to the GV model, whose ratio to $Z_0$ imparts a different behavior to each member of GVSC. Linear dimensions of the plasma, which scale with anode radius a, and its density, which scales with the gas density $\rho_0$, are related to parameters of the capacitor bank and dimensionless parameters of the GV model as:

$$a = 2\pi L_0 \mu_0^{-1} \kappa = 5(\text{mm}) L_0(\text{nH}) \kappa \; ; \qquad 16$$

$$\rho_0 = \left(\frac{\mu_0}{2\pi}\right)^5 \left(\frac{V_0^2 C_0^2}{L_0^4}\right)\left(\frac{\varepsilon^2}{\pi\kappa^4}\right)$$
$$= 1.018\times 10^{-4} (\text{kg}/\text{m}^3) \left(\frac{V_0^2(\text{kV}) C_0^2(\mu F)}{L_0^4(\text{nH})}\right)\left(\frac{\varepsilon^2}{\kappa^4}\right) \qquad 17$$

The mass density of gas of molecular weight A at normal temperature is related to its pressure as

$$\rho_0(\text{kg}/\text{m}^3) = 4.464\times 10^{-2} A \cdot p(\text{bar}) \qquad 18$$

The quantity $p_{CH}$

$$p_{gas} \equiv p_{CH}\left(\frac{\varepsilon^2}{\kappa^4}\right); p_{CH} \equiv \frac{2.28}{A}(mbar)\left(\frac{V_0^2(kV)C_0^2(\mu F)}{L_0^4(nH)}\right) \quad\quad 19$$

is a characteristic pressure scale for a gas of molecular weight A for a given capacitor bank.

Empirical observations [17] that the pinch radius ~0.12a and pinch height ~0.8a can be used to estimate the *scaling* of mass content of the pinch zone (not taking into account plasma compression) as

$$M_{pinch} \sim \pi(0.12a)^2(0.8a)\rho_0$$
$$= 0.011\left(\frac{\mu_0}{2\pi}\right)^2\left(\frac{V_0^2 C_0^2}{L_0}\right)\left(\frac{\varepsilon^2}{\pi\kappa}\right) \quad\quad 20$$
$$= 1.47\times 10^{-13}(kg)\left(\frac{V_0^2(kV)C_0^2(\mu F)}{L_0(nH)}\right)\left(\frac{\varepsilon^2}{\kappa}\right)$$

The scale of axial and radial velocities is given by

$$v_A = \frac{\mu_0 I(t)}{\pi a\sqrt{2\mu_0\rho_0}} = \mathbb{V}\frac{2\pi}{\mu_0}\sqrt{\frac{L_0}{C_0}} = 1.58\times 10^5 (m/sec)\mathbb{V}\sqrt{\frac{L_0(nH)}{C_0(\mu F)}} \text{ for } D_2; \mathbb{V} \equiv \kappa\varepsilon^{-1}\tilde{I}(\tau) \quad\quad 21$$

This equation reveals a characteristic velocity scale, numerically related to the drive parameter by a multiplicative constant, which, for every capacitor bank, is proportional to its impedance. The scales of linear dimension and of plasma velocity define the scale of time for gross plasma motion:

$$t_{scale} \sim a/v_A \sim \frac{\varepsilon}{\tilde{I}(\tau)}\sqrt{L_0 C_0} \sim 3.16\times 10^{-8}(sec)\sqrt{L_0(nH)C_0(\mu F)}\frac{\varepsilon}{\tilde{I}(\tau)} \quad\quad 22$$

X-ray and neutron yields would then scale as the yield parameter $\mathbb{Y}$,

$$\mathbb{Y} \equiv \rho_0^2 a^3 t_{scale} = 6.58\times 10^{-42}(kg^2\text{-}sec/m^3)\mathbb{Y}_0\frac{V_0^6(kV)C_0^{6.5}(\mu F)}{L_0^{8.5}(nH)};$$
$$\mathbb{Y}_0 \equiv \left(\frac{\varepsilon^7}{\kappa^9\tilde{I}(\tau_R)}\right) \quad\quad 23$$

The intensity and spectral properties of the x-ray and neutron emission would depend on the details of the plasma processes in the pinch phase – an important aspect, which is beyond the scope of the GV model.

Typically, the cost of a capacitor bank for the DPF facility should scale as its energy. One could define a cost-effectiveness parameter (CEP) in terms of 23:

$$\text{CEP} \equiv \frac{\rho_0^2 a^3 t_{scale}}{\tfrac{1}{2} C_0 V_0^2} = 1.3 \times 10^{-41} \left(\text{kg}^2 \text{ sec}/\text{m}^3 \text{J}\right) \mathbb{Y}_0 \frac{V_0^4 (\text{kV}) C_0^{5.5} (\mu\text{F})}{L_0^{8.5} (\text{nH})} \qquad 24$$

This indicates that within a GV similarity class (GVSC), it is far more cost effective to choose a lower inductance than to increase the voltage and capacitance as far as the yields of emissions from the DPF are concerned.

Sometimes, yield may not be the only desired goal; one may want a single, short, intense pulse of neutrons as a diagnostic probe. From 22, the optimization goal may include a smaller value of $\varepsilon / \tilde{I}(\tau_R)$ and members of the GVSC with lower $\sqrt{L_0 C_0}$ may be preferred.

An important caveat needs to be mentioned here. The GV model is a theory of the consequences of the snowplow hypothesis in the context of the DPF geometry: *it is not a theory of any of the plasma phenomena*. It neglects the details of the plasma formation process at the insulator [39,40] as well as of the processes [41] which lead to the approximate validity of the snowplow hypothesis. It also neglects plasma processes which govern the temperature and energetic-ion velocity distribution in the pinch phase. It is seen from 17 that every member of the similarity class of the GV model is associated with a characteristic scale of the fill density. For some values of the fill density, the plasma processes associated with the formation phase and /or leading to the snowplow phenomenon may not proceed in an optimal manner in the conventional design of a DPF. These are considerations which lie beyond the scope of the present formulation of the GV model and represent important subjects of research and innovation in their own right.

## 4. Quantitative performance criteria and optimal properties of GVSC:

From the point of view of device optimization, the snowplow phenomenon represented in the GV model using a set of dimensionless parameters may be looked upon as just a mechanism for delivering energy stored in the capacitor bank to a dense plasma formation process *at the end of the rundown phase* of a Mather type DPF; this plasma formation process may proceed in somewhat different manner for devices having different shapes for the anode end-caps. The objective of a theoretical optimization procedure is to provide the initial values of parameters for an iterative *empirical* optimization campaign to assist its rapid convergence; *theoretical optimization has no meaning other than as a prelude to such indispensable empirical optimization*. Because of this, theoretical optimization needs to consider only the zero-resistance case; for sufficiently low values of the circuit resistance ($\gamma<1$), the zero-resistance optimum configuration should provide a good starting point for the iterative empirical optimization. In practice, the scaled radius of the insulator $\tilde{r}_I$ does not change much, and may be taken as nearly equal to 1, effectively reducing the domain of optimization to the 5-dimensional parameter space consisting of the parameters $\varepsilon, \kappa, \tilde{z}_A, \tilde{z}_I, \tilde{r}_C$. This 5-D parameter space has natural boundaries defined in the GV model by the requirement that both $\eta_m$ and $\eta_W$ be bounded between 0 and 1. For example, it is clear from 9 that $\varepsilon(\tau_R +1) < 2$ to ensure $\eta_W > 0$ throughout the DPF evolution for the zero-resistance case. Within such bounded 5-D parameter space, a "region of practical interest" can be identified from accumulated worldwide research experience.

The optimization problem involves choosing the snowplow device parameters to yield the 'best results' for a given capacitor bank. The very idea of optimization implies existence of opposing trends in desired quantitative performance criteria; the first step in global optimization of DPF must therefore be identification of desirable quantitative performance goals and of opposing tendencies. Study of optimal properties of GV model is facilitated by defining the following numbers [30], also bounded between 0 and 1:

$$\mu^0 \equiv 1 - m_0(\tau_R)/\tau_R \mathcal{L}(\tau_R), \quad \mu^1 \equiv 1 - m_1(\tau_R)/\tfrac{1}{2}\tau_R^2 \mathcal{L}(\tau_R), \quad X \equiv \kappa\mathcal{L}(\tau_R)/(1+\kappa\mathcal{L}(\tau_R)).$$

This yields the expressions

$$\eta_M = 2\varepsilon\tau_R (1-\mu^0 X) - \varepsilon^2 \tau_R^2 (1-\mu^1 X) \qquad 25$$

$$\eta_W = X\varepsilon\tau_R (2\mu^0 - \varepsilon\tau_R \mu^1) \qquad 26$$

One of the possible quantitative performance criteria is the fraction of energy converted into magnetic energy coupled *with plasma inductance* at the end of rundown phase [29]:

$$\eta_{MP}(\tau_R) = \kappa\mathcal{L}(\tau_R)\tilde{I}^2(\tau_R) = 2\varepsilon\tau_R X(1-\mu^0 X) - \varepsilon^2 \tau_R^2 X(1-\mu^1 X) \qquad 27$$

In applications involving material modification, the total energy (magnetic energy + work done) coupled with plasma inductance may be more important than only the magnetic energy:

$$\eta_T(\tau_R) \equiv \eta_{MP}(\tau_R) + \eta_W(\tau_R)$$
$$= 2\varepsilon\tau_R X(1+\mu^0(1-X)) - \varepsilon^2 \tau_R^2 X(1+\mu^1(1-X)) \qquad 28$$

It is easily seen that these performance criteria have a maximum with respect to $\varepsilon\tau_R$, which involves the gas pressure, the anode radius, the charge on the capacitor bank and the scaled lengths of anode and insulator. But the maximum value asymptotically reaches unity as a function of $\kappa$, which represents the physical size of the DPF for a given capacitor bank. This is understandable, since as the size increases, the DPF inductance becomes the dominant inductance in the circuit and hence will acquire most of the capacitor energy at the peak of current in a resistance-less circuit. The 'penalty' for indiscriminate increase in physical size would be increase in the discharge time, which is not reflected in any of the expressions 25-28.

This suggests that although the definition of 'best results' may differ from application to application, a common desirable optimal feature would be maximum conversion of stored capacitor bank energy into magnetic energy associated with plasma inductance *in minimum time* so that parasitic energy losses from radiation, heat conduction to electrodes and dissipation in circuit resistance are minimized. This criterion is applied to the end of the rundown phase because that is the dominant phase in the evolution of the Mather type plasma focus, because the anode top is many times configured to have a cavity or a hemispherical or conical profile, which is not taken into account in the formula 12 for the dynamic inductance and also because the GV model becomes less accurate near the pinch phase in view of its neglect of gas dynamics which plays a significant role at that stage. It was reported [29] earlier that for given values of $\tilde{z}_A, \tilde{r}_C, \tilde{r}_I$ and $\tilde{z}_I$, the average power parameter (APP) $\mathbb{P}$ defined as

$$\mathbb{P} \equiv \eta_{MP}(\tau_R) / \tilde{t}(\tau_R) \qquad\qquad 29$$

has a well-defined maximum in $(\kappa, \varepsilon)$ space for the facilities PF-1000 and Lawrenceville Plasma Physics (LPP). The occurrence of maximum with respect to $\varepsilon$ is already mentioned earlier; the maximum with respect to $\kappa$ comes because both $\eta_{MP}(\tau_R)$ and $\tilde{t}(\tau_R)$ increase at different rates as the size of the DPF increases.

It is reasonable to assume that a globally optimized DPF will be a member of the set of all DPF facilities which maximize the average power transferred to the plasma during the rundown phase. The problem of maximizing the average power parameter $\mathbb{P}$ is analogous to the maximum power transfer theorem of electrical circuit theory, which leads to an impedance matching condition between the power source and the electrical load. The existence of a maximum value of $\mathbb{P}$ as a function of $\kappa$ similarly translates into a relation between the static inductance of the power source (capacitor bank) and the plasma inductance at the end of rundown. It leads to the conclusion that the physical size of the device which maximizes the average power transfer within a GVSC should be related to the static inductance of the circuit *and should be independent of the voltage*. This point is revisited in section V.

It appears from 21 that the plasma velocities should unconditionally scale with the capacitor bank impedance [1]. However, there is an important consideration, first pointed out by Bruzzone, Kelly, Milanese and Pouzo (BKMP) [42] and subsequently elaborated in other publications [43,44,45,46], that the electromagnetic work done in accelerating the PCS must be adequate to heat, dissociate and ionize the entire mass of gas swept up. The energy available for this purpose is [43] one-half of the electromagnetic work done; the remaining half is the kinetic energy of the plasma. This criterion is easily formulated in terms of the GV model using 9, 14, and 17

$$\tfrac{1}{2} W_W / \text{swept\_mass} = \left(\frac{2\pi}{\mu_0}\right)^2 \frac{L_0}{C_0} \mathbb{S} \; (J/kg) \geq \mathbb{E}_S \qquad\qquad 30$$

The quantity $\mathbb{E}_S$ in 30 is the specific energy $(\sim 1.24 \text{MJ}/\text{gm})$ necessary to heat (upto ~10 eV), dissociate [47] and ionize deuterium gas initially at room temperature; $\mathbb{S}$, the 'specific energy parameter', is defined by

$$\mathbb{S} \equiv \frac{1}{\left(\tilde{r}_c^2 - 1\right)\left(\tilde{z}_A - \tfrac{3}{7}h\right)} \frac{\kappa \eta_w}{\varepsilon^2} \qquad 31$$

It is interesting to note that 30 introduces a characteristic impedance associated with the deuterium gas

$$Z_D \equiv \left(\frac{\mu_0}{2\pi}\right)\sqrt{\mathbb{E}_S} \approx 7.04 \text{ m}\Omega \quad \left(\text{for } \mathbb{E}_S \sim 1.24 \text{MJ}/\text{gm}\right) \qquad 32$$

Each capacitor bank is then characterized by an additional dimensionless number

$$\mathbb{B}_0 \equiv \frac{Z_D}{Z_0} = \frac{\mu_0}{2\pi}\sqrt{\frac{\mathbb{E}_S C_0}{L_0}} \qquad 33$$

The criterion 23 can be stated as an equality in terms of a "BKMP factor" $f_B$ as

$$\mathbb{S} = f_B \mathbb{B}_0^2; f_B \geq 1 \qquad 34$$

This factor is the ratio of the left and right hand sides of the inequality 30. It should be "sufficiently" larger than unity to accommodate the presently inexact numerical definition of $\mathbb{E}_S$, which depends on the change in thermodynamic state of the gas. Combination of 21 and 30 leads to the relation

$$v_A = \mathbb{Q}\sqrt{f_B \mathbb{E}_S}; \quad \mathbb{Q} \equiv \mathbb{V}\mathbb{S}^{-0.5} \qquad 35$$

The striking feature of 35 is the existence of a lower limit (in terms of the BKMP factor $f_B$) on the scale of velocity, whose numerical value depends on the square root of $\mathbb{E}_S$ $\left(\sim 3.520 \times 10^4 \text{ m/s for } \mathbb{E}_S \sim 1.24 \text{MJ}/\text{gm}\right)$, which is a physical property of the working gas, and dimensionless quantities related to the GV model and *which does not additionally depend on any property of the capacitor bank including its energy.*

In the neighborhood of an optimal point (using any criterion) of the GV model, the quantities $\mathbb{Q}$ and $\mathbb{S}$ are seen to vary much more slowly than $\mathbb{Y}_0$, the reduced yield parameter. This probably explains the empirically observed [17] near-constancy of the drive parameter (proportional to $v_A$) in neutron-optimized DPF devices over 8 decades in capacitor bank energy. Relation 21 also suggests existence of an upper limit on $v_A$ related to the capacitor-bank-dependent characteristic velocity $2\pi\mu_0^{-1}\sqrt{L_0 C_0^{-1}}$ and GV model bounds on the factor $\mathbb{V} \equiv \kappa\varepsilon^{-1}\tilde{I}(\tau)$ for an optimal configuration. One could increase the velocity parameter $\mathbb{V}$ beyond the GV model upper bound only at the cost of decreased transfer of energy to the plasma; this is the point made by Klir and Soto [1].

The BKMP criterion is a global formulation of the condition that the energy required to maintain the plasma state must come out of the work done by the electromagnetic force. It should in principle be possible to formulate this condition in terms of local values of parameters. However, such local formulation is beyond the scope of this paper, which revolves around the GV model; it forms the subject matter of a forthcoming paper (under review).

Cardenas [7] has proposed maximization of magnetic energy per particle in the pinch phase as a design criterion. This can be formulated in the following manner. The fraction of energy converted into magnetic energy at the end of radial implosion phase can be estimated as

$$\eta_{MP}(\tau^*) = \kappa\mathcal{L}(\tau^*)\tilde{I}^2(\tau^*) \qquad 36$$

Using 20, the magnetic energy per particle $\mathcal{E}$ for plasma of average ion mass $m_i$ and effective atomic number $Z_{eff}$ should then scale as

$$\begin{aligned}
\mathcal{E} &\equiv \frac{1}{2}C_0 V_0^2 \cdot \eta_{MP}(\tau^*) \bigg/ \left((Z_{eff}+1)\cdot M_{pinch}/m_i\right) \\
&= \frac{142.8}{(Z_{eff}+1)}\left(\frac{2\pi}{\mu_0}\right)^2 \frac{L_0}{C_0} m_i \mathcal{E}_{GV}; \quad \mathcal{E}_{GV} \equiv \frac{\kappa^2}{\varepsilon^2}\mathcal{L}(\tau^*)\tilde{I}^2(\tau^*) \qquad 37 \\
&\approx 38(eV)\frac{L_0(nH)}{C_0(\mu F)}\mathcal{E}_{GV} \text{ for } D_2
\end{aligned}$$

The energy per particle is also seen to be related with the capacitor-bank-dependent characteristic velocity $2\pi\mu_0^{-1}\sqrt{L_0 C_0^{-1}}$. Relation 37 does not take into account plasma compression; the actual value of energy per particle would be smaller by the compression ratio. The value of this parameter works out to 66 eV for PF-1000 [29]. Because of the progressive inaccuracy of the GV model near the pinch phase pointed out earlier, this cannot be used as a primary optimality criterion in the present version of the GV model; however, it may be used as a secondary optimality criterion to rank configurations shortlisted from a primary optimization scheme. The same consideration applies to the use of the pinch current $\tilde{I}(\tau^*)$ at the limit of validity of the GV model as an optimality criterion.

The above discussion illustrates the value of the idea of GV model similarity class. Using a comparative study of devices which have the same representation in GV model but differing physical parameters, one can look at deviations from predictions of the resistive GV model, which should contain signatures of phenomena which violate the conditions of proper plasma formation assumed in the resistive GV model. Understanding conditions of proper plasma formation is one of the major goals of plasma focus research and the concept of the GVSC proposed in this paper provides a tool to secure a major advance in that direction.

5. **Parameter space survey of GVSC:**

Optimization search algorithms are a research subject by themselves; however, application of such algorithms to the determination of a globally optimized DPF requires a prior determination of an adequate definition of optimality. This is difficult for the case of DPF because of a profusion of diverse end-applications and practical considerations of an industrial-scale investment already mentioned earlier. The next best option is to perform unbiased tabulation of optimality parameters in a uniformly discretized parameter sub-space as a permanent database, on which to perform optimization search tailored to specific end-applications. Different optimization queries on this database are *expected* to come up with different answers; the design of the query would sensitively depend on the type of application. This section therefore attempts to provide illustrative examples of two variations of optimization search and to highlight some counter-intuitive aspects of the optimization process.

The formulation of GV model in terms of dimensionless parameters allows a universal determination of its optimal properties, valid for all members of a GVSC, justifying more extensive efforts than would be practical for optimization of DPF for a particular capacitor bank (such as PF-1000). This has taken the form of once-for-all generation of a database of properties associated with very large number of points in the 5-dimensional parameter sub-space of the zero-resistance GV model, chosen in the following manner.

Many working devices (notably PF-1000, used as an example in this paper) use the scaled insulator length $\tilde{z}_I$ close to 1; this was chosen as a fixed value in this series of investigations although there are indications that the optimum value may be different. The scaled cathode radius $\tilde{r}_C$ was varied from 1.2 to 2.0 in steps of 0.1. For each value of $\tilde{r}_C$, the scaled anode length $\tilde{z}_A$ was varied from $\mathrm{Max}\left[2, \tilde{r}_C, 0.5\tilde{r}_C^2 - 1 + \tilde{r}_I\right]$ to 10.0 in steps of 0.1. For each value of $\tilde{z}_A$, the value of $\epsilon$ was varied from $0.1\tau_R$ to $1.99\tau_R$ in steps of $0.1\tau_R$. For each value of $\epsilon$, the value of $\kappa$ was varied from 0.1 to 2.0 in steps of 0.1. At each point, the set $\left\{\tilde{z}_I, \tilde{z}_A, \tilde{r}_c, \kappa, \epsilon, \mathbb{P}, \eta_M, \eta_{MP}, \eta_W, \eta_T, \tilde{I}(\tau_R), \mathbb{Q}, \mathbb{S}, \mathbb{Y}_0, \mathbb{V}, \int_0^{\tau_R} \tilde{I}(\tau)d\tau, \mu_0, \mu_1\right\}$ was calculated and saved in a database.

For DPF as a source of radiation, the "performance" has many aspects: it must have the best power transfer efficiency, highest current, highest conversion into magnetic energy of the plasma as well as high cost-effectiveness parameter CEP; for a GVSC, this translates to best $\mathbb{Y}_0$. On the other hand, some applications of DPF (such as material modification) may prefer to maximize the average total power parameter $\mathbb{P}_T \equiv \eta_T(\tau_R)/\tilde{t}(\tau_R)$ along with high plasma velocity parameter $\mathbb{V}$ and a high fill pressure of gas. These two cases were studied separately *using the same database* to provide comparison and contrast.

For the first case, the database for the chosen value of $\tilde{z}_I = 1.0$, (containing 1,45,800 cases calculated over 10 days) was searched for the maximum value of $\mathbb{P}$ and the points having values of $\mathbb{P}$ between 90% and 100% of the maximum were shortlisted. This short list (SL-Ia) containing 11683 cases, was arranged in the order of decreasing values of the scaled current $\tilde{I}(\tau_R)$ and cases lying between 90% and 100% of maximum value of $\tilde{I}(\tau_R)$ were shortlisted in the second shortlist (SL-IIa) containing 549 configurations. This was arranged according to decreasing values of $\eta_{MP}$ and those cases lying between 90% and 100% of the maximum value of $\eta_{MP}$ were taken into a shortlist III (SL-IIIa), containing 147 cases. This shortlist was arranged in the order of decreasing $\mathbb{Y}_0$ and cases lying between 50% and 100% of maximum value of $\mathbb{Y}_0$ were put in shortlist IVa (SL-IVa) containing 7 cases. This was arranged in order of decreasing values of the parameter $\mathbb{V}$ in 21 and cases lying between 90% and 100% of its maximum value were taken into the fifth shortlist containing 4 semi-finalists. These 4 cases (see Table I) then represent the best combination of highest average power transfer to the plasma, highest scaled current, most efficient magnetic energy transfer to the plasma, best yield parameter and maximum plasma velocity for a given capacitor bank. They are quite similar in their properties so only the first case, with the highest value of the pressure parameter $\varepsilon^2/\kappa^4$, was chosen as Global Optimum I within the context of this study, subject to various caveats and conditions already mentioned.

The second optimization search arranged the *same database* in decreasing values of $\mathbb{P}_T = \mathbb{P} \cdot \eta_T/\eta_{MP}$ and cases lying between 95% and 100% of its maximum value were shortlisted into SL-Ib containing 2009 configurations. This shortlist was ranked in decreasing values of $\mathbb{V}$ and cases lying between 95% and 100% of its maximum value were taken into SL-IIb containing 18 cases. This was ranked according to decreasing values of the pressure parameter $\varepsilon^2/\kappa^4$ and cases lying between 80% and 100% of its maximum value were taken into the final shortlist containing 4 almost similar configurations shown in Table-II. The configuration with the highest velocity parameter $\mathbb{V}$ was designated as Global Optimum II.

Note that these values, including specific optimal values of $\kappa$, are independent of the nature of the gas and of the capacitor bank voltage, capacitance and inductance. *This shows that the optimal size of the DPF, represented by anode radius a, is dependent on the capacitor bank inductance and independent of its voltage.*

Table I: Properties of Global Optimum I

| # | $\tilde{z}_A$ | $\tilde{r}_C$ | $\kappa$ | $\varepsilon$ | $\mathbb{P}$ | $\eta_{mp}$ | $\eta_w$ | $\tilde{I}(\tau_R)$ | $\mathbb{Y}_0 \times 10^5$ | $\int_0^{\tau_R} \tilde{I}(\tau) d\tau$ | $\mathbb{Q}$ | $\mathbb{S}$ | $\mathbb{V}$ | $10^2 \cdot \varepsilon^2 / \kappa^4$ |
|---|---|---|---|---|---|---|---|---|---|---|---|---|---|---|
| 1 | 3.1 | 1.5 | 0.9 | 0.190 | 0.326 | 0.370 | 0.246 | 0.585 | 3.88 | 2.12 | 1.79 | 2.40 | 2.78 | 5.47 |
| 2 | 2.8 | 1.5 | 1.0 | 0.221 | 0.326 | 0.371 | 0.245 | 0.586 | 4.41 | 1.95 | 1.90 | 1.95 | 2.65 | 4.89 |
| 3 | 2.2 | 1.4 | 1.4 | 0.332 | 0.325 | 0.366 | 0.222 | 0.609 | 3.51 | 1.42 | 2.16 | 1.42 | 2.57 | 2.86 |
| 4 | 2.1 | 1.4 | 1.5 | 0.362 | 0.325 | 0.368 | 0.226 | 0.604 | 3.50 | 1.30 | 2.20 | 1.30 | 2.50 | 2.59 |

The scaled current profile for Global Optimum I is shown in Fig 1 for $\gamma=0$ and 0.2. For the zero resistance case, the partitioning of energy at the end of rundown is $\eta_M = 0.711$, ($\eta_{MP} = 0.37$), $\eta_W = 0.247$, $\eta_C = 0.042$, $\mathcal{E}_{GV} = 8.82$. For $\gamma=0.2$, $\eta_M = 0.587$, ($\eta_{MP} = 0.306$), $\eta_W = 0.296$, $\eta_C = 0.042$, $\eta_R = 0.075$, $\mathcal{E}_{GV} = 6.91$. The GV model fit to PF-1000 data reported earlier [29] is included for comparison.

Table II: Properties of Global Optimum II

| # | $\tilde{z}_A$ | $\tilde{r}_C$ | $\kappa$ | $\varepsilon \times 100$ | $\mathbb{P}_T$ | $\eta_{mp}$ | $\eta_w$ | $\tilde{I}(\tau_R)$ | $\mathbb{Y}_0 \times 10^{11}$ | $\int_0^{\tau_R} \tilde{I}(\tau) d\tau$ | $\mathbb{Q}$ | $\mathbb{S}$ | $\mathbb{V}$ | $10^4 \cdot \varepsilon^2 / \kappa^4$ |
|---|---|---|---|---|---|---|---|---|---|---|---|---|---|---|
| 1 | 9.5 | 1.2 | 1.3 | 3.51 | 0.599 | 0.373 | 0.299 | 0.408 | 1.52 | 6.40 | 0.82 | 341.3 | 15.1 | 4.32 |
| 2 | 9.4 | 1.2 | 1.3 | 3.55 | 0.599 | 0.372 | 0.297 | 0.410 | 1.65 | 6.35 | 0.82 | 331.9 | 15.0 | 4.42 |
| 3 | 10 | 1.2 | 1.2 | 3.32 | 0.601 | 0.370 | 0.298 | 0.412 | 2.07 | 6.84 | 0.80 | 351.9 | 14.9 | 5.30 |
| 4 | 9.3 | 1.2 | 1.3 | 3.60 | 0.600 | 0.372 | 0.296 | 0.412 | 1.78 | 6.29 | 0.83 | 322.6 | 14.9 | 4.53 |

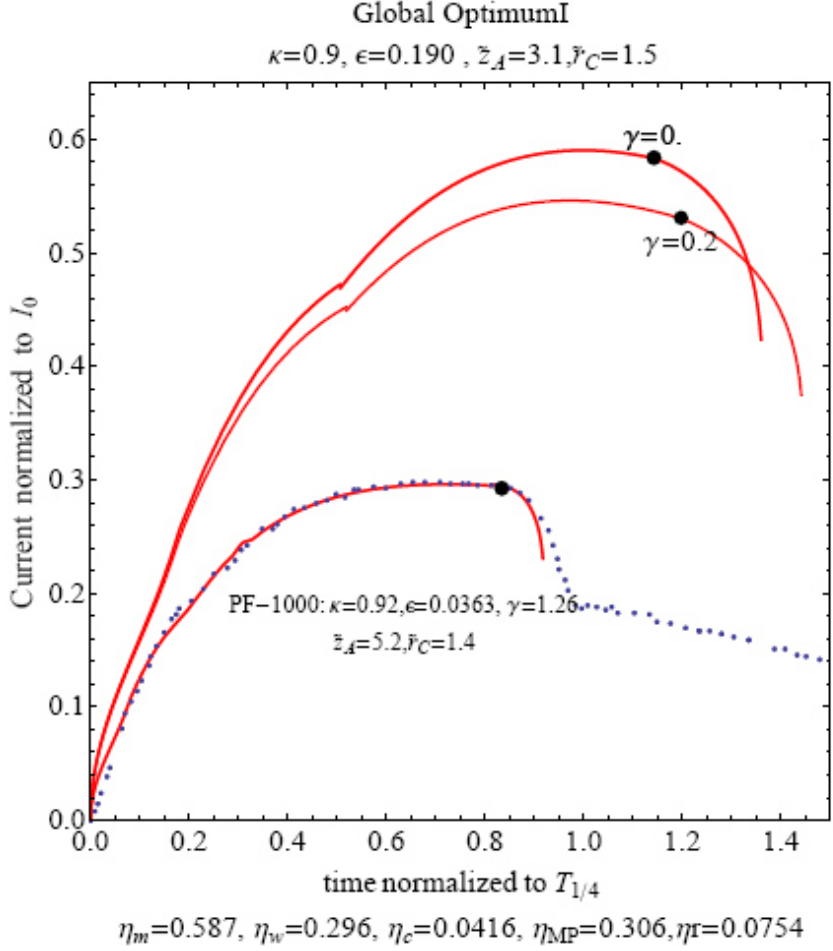

Figure-2: Scaled current profiles for the Global Optimum I for $\gamma = 0$ and $0.2$. GV model fit to PF-1000 reported earlier [29] is included for comparison. The black dot represents the end of rundown phase. Beyond the rundown phase, the GV model becomes progressively less accurate as neglected gas dynamic phenomena start becoming important during radial implosion.

The upper limit of velocity scale in 21 is given by $\mathbb{V} \leq 2.78$; the value of $\mathbb{Q}$ in Table I gives $v_A \sim 1 \times 10^5$ m/s for $f_B \sim 2.5$. In the original database, the maximum and minimum values of $\mathbb{V}$ are 40 and $1 \times 10^{-5}$ and those of $\mathbb{Q}$ are 3.26 and $1.4 \times 10^{-4}$. The optimization process described above thus leads to selection of minimum and maximum values of the velocity scale (proportional to the drive parameter) irrespective of the scale of energy [17]. Attempts to increase the velocity beyond this limit by decreasing pressure would result in poor energy transfer to the plasma.

The scaled current profile for Global Optimum II is shown in Fig 3 for $\gamma=0$ and $0.2$. For the zero resistance case, the partitioning of energy at the end of rundown is $\eta_M = 0.539$, ($\eta_{MP} = 0.373$), $\eta_W = 0.298$, $\eta_C = 0.162$, $\mathcal{E}_{GV} = 348$. For $\gamma=0.2$, $\eta_M = 0.478$, ($\eta_{MP} = 0.330$), $\eta_W = 0.317$, $\eta_C = 0.162$, $\eta_R = 0.043$, $\mathcal{E}_{GV} = 307.1$.

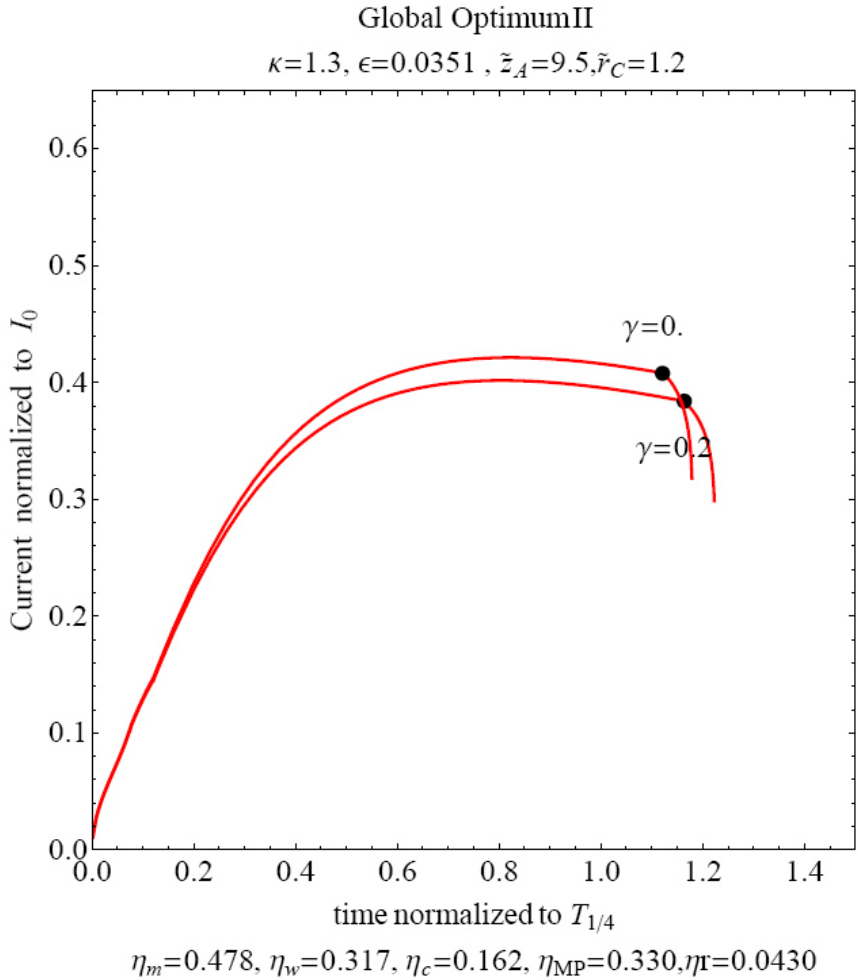

Figure-3: Scaled current profiles for the Global Optimum II for $\gamma = 0$ and $0.2$. The black dot represents the end of rundown phase.

$\eta_m=0.478$, $\eta_w=0.317$, $\eta_c=0.162$, $\eta_{MP}=0.330$, $\eta_T=0.0430$

The two configurations have differences as well as similarities: the first configuration has more peak current, but both configurations have similar values of $\eta_{MP}$. The 45% higher current of the first configuration comes from a better utilization of stored energy: only 4% energy remains in the capacitor bank at the end of rundown as compared with 16.2% for the second configuration. The energy per particle parameter $\mathcal{E}_{GV}$ is 40 times higher and the velocity parameter $\mathbb{V}$ is 5.4 times higher in the second configuration; the pressure parameter is however 127 times lower and the yield parameter is $2.6 \times 10^6$ times smaller. In terms of device geometry, the first configuration has 1.44 times smaller anode radius and 3 times smaller anode length. This exercise shows the profound effect of the nature of the query on the outcome of the optimization search; it also shows that empirical and theoretical optimization procedures have a mutually supportive complementary role to play where theory provides a wide coverage of the parameter space while experiments provide validation to specific conclusions of the theoretical optimum search.

It needs to be remembered that the GV model cannot be used to make any predictive statements concerning the neutron yield; therefore *the above configurations do not necessarily represent neutron-optimized regimes*. However, it should not be surprising to *experimentally discover* that they do indeed have enhanced neutron emission properties in view of the high current and high magnetic energy.

The above global optimization scheme could in principle be extended in future by incorporating additional physics external to the GV model which would further restrict the searchable parameter subspace, leading to more efficient optimization. Two such examples can be mentioned. It may be possible to incorporate some elements of ionization dynamics in the form of a generalization of Paschen's Law or ionization stability condition. Another could be incorporation of a model of nuclear reactions from fast ions interacting with a target plasma within which they remain confined. Both these aspects require considerable preparatory work before they can be taken into account in the global optimization.

The following example illustrates how a globally-optimized DPF facility with stored energy of 1 MJ and design short-circuit current $I_0=10$ MA (static inductance 20 nH) might be realized. Using 34, with $f_B = 2$ (a conservative safety factor) and $\mathbb{S} = 2.4$ from Table-I, $Z_0 = 6.42$ m$\Omega$. The operating voltage should therefore be 64.2 kV and the capacitance should be 485 $\mu$F. Using $\kappa=0.9$ from Table-I, the anode radius becomes 90 mm and the insulator length is then also 90 mm. The anode length and the cathode inner radius become 279 mm and 135 mm respectively. From 19 and the value of the pressure parameter from Table I, the deuterium fill pressure comes to 187.75 mbar. *Note that the BKMP criterion is well-satisfied* even though the value of pressure is much higher than that of any operating DPF. This provides a counter-example to the suggestion that DPF devices have an inherent limitation on their operating pressure related to the BKMP criterion [43]. At a circuit resistance of 1.28 m$\Omega$ ($\gamma$=0.2), the magnetic energy in the dynamic inductance at the end of the rundown should be 300 kJ. The current at the end of rundown should be ~5.3 MA, the pinch current should be more than 3.5 MA.

This may be compared with an extreme example: a hypothetical globally optimized facility with static inductance of 20 nH and $I_0$=100 MA, needing 100 MJ of stored energy. The impedance of the capacitor bank is still $Z_0 = 6.42$ m$\Omega$ according to the logic described in the previous paragraph and the capacitance remains 485 $\mu$F; the operating voltage however becomes 10 times higher at 642 kV. *The anode radius, insulator length, anode length and cathode inner radius remain the same as for the 10 MA case.* But the pressure of deuterium works out to 345 bar! At a circuit resistance of 1.28 m$\Omega$ ($\gamma$=0.2), the magnetic energy in the dynamic inductance at the end of the rundown should be 30 MJ; about 7.5 MJ should be dissipated in the circuit resistance. The current at the end of rundown should be ~53 MA and pinch current should be more than 35 MA. These are impressive technological challenges but *they do not represent limitations imposed by the physics of primary energy transport via the snowplow effect in DPF* as represented by the GV model [48]. The conclusion [48] that the large capacitor banks cannot drive Mather type DPF devices to multi-mega-amperes pertains to a design procedure that keeps the voltage rather than the impedance constant as pointed out above.

This example highlights the following counterintuitive aspect of optimization within the ambit of the RGV model. The optimization happens in the dimensionless model parameter space, in terms of quantitative performance criteria defined from first principles, with no reference to empirical thumb-rules, *without regard for practicalities of realization of the DPF configuration, which reflect the accessible sophistication of device technology*. On the one hand, this presents the best expected performance under well-defined conditions, which cannot be exceeded even in principle, thus affording protection from the risk of premature technical obsolescence. On the other hand, it allows customization of the search algorithm to avoid bias rooted in existing state-of-art facilitating identification of areas where investment in innovation can reap rich dividends in terms of clear technical advantages over existing state-of-art. This feature recognizes the distinction between a widely practiced consensus and limitations imposed by laws of nature. *This needs to be seen in the context of applicability of the RGV model for arbitrary devices based on the snowplow effect, not limited to the Mather type DPF as in this study.*

6. **Summary and conclusions:**

This paper provides a first look at the possibility of theoretical global optimization, as the first step of an empirical optimization campaign, of a Mather type DPF using a formula for dynamic inductance reported earlier [29] representing a numerical fit to thousands of automated calculations of 2-D plasma profile provided by the GV model. Feasibility of such global optimization is an essential prerequisite to the emergence of DPF as a technology platform for diverse commercial applications, from plasma nanotechnology and material modification [24,25] to use of intense neutron bursts [26,27] and production of short-lived radioisotopes [28] for medical applications to fusion energy [23] using advanced fuels. The desirability of such optimization can be inferred from the fit of PF-1000 current profile to the resistive GV model (Fig. 1) which shows about 47.5% energy remaining in the capacitor bank and 10% dissipated in circuit resistance by the end of rundown - a clearly unsuitable situation in a commercial context.

The utility of the GV model for this purpose lies in its dimensionless form, which maps 10 physical parameters representing a DPF installation on to 7 independent dimensionless model parameters. Each point in the GV model parameter space then has three degrees of freedom in the space of all possible DPF configurations, which are chosen as the voltage, capacitance and inductance of the capacitor bank. The present discussion of optimization in a 4-dimensional parametric subspace of the GV model includes the BKMP criterion which requires the electromechanical work done in plasma propagation to exceed an energy threshold related to the change of the thermodynamic state of the working gas from an initially neutral gas at room temperature to fully-dissociated, ionized and sufficiently heated plasma. This reveals a characteristic impedance associated with the working gas, whose ratio to the characteristic impedance of the capacitor bank is a dimensionless number characteristic of a capacitor bank. Together with the ratio of the static resistance of the capacitor bank to its impedance, this ratio should form part of the definition of a *generic technological limit* on the maximum performance of a Mather type DPF [48]; the importance of the present approach is that it allows a precise demarcation of this limit at a given level of capacitor bank energy, potentially leading to innovative workarounds.

The scope of this study can clearly be expanded to the discovery of optimum configurations for a specific capacitor bank, such as PF-1000. The GV model fit to current waveform from PF-1000 reveals a high value of circuit resistance compared to circuit impedance ($\gamma$=1.26). The strategy of using optimum configurations of the zero-resistance case as initial step of optimization for low $\gamma$ situations then becomes of doubtful validity. This suggests generation of a database specific to PF-1000. The optimum value of $\kappa$ (0.9) for the Global Optimum I configuration is seen to be quite close to that of PF-1000 (0.92) as revealed through the fitted value of $L_0$=25 nH; the value of $\tilde{z}_I$ for PF-1000 is 0.98, quite close to that used for the database calculation. Therefore, the existing anode of radius 115 mm, and existing insulator (of radius 115 mm and length 113 mm) of PF-1000 device can be retained and the procedure outlined above can be used to determine the anode length, cathode diameter and pressure that would maximize average power transfer, current and energy per particle. This however requires an iterative solution to 6 for each point in the database, which is computationally much more resource intensive and has to be a separate undertaking forming part of a project for upgradation of existing large facilities.

This study also suggests that several physical phenomena, which are responsible for the approximate validity of the snowplow effect and which probably play an important role in limiting the pressure range for neutron producing devices, need to be incorporated in a future extension of the GV model. The GV model should therefore prove to be a rewarding subject of both theoretical and experimental research.

Acknowledgements: The author would like to thank Prof. H. Bruzzone for making available the dissertation of J. M. Vargas and Dr. M. N. Giridhar Gopal of Scientific Information Resource Division, Bhabha Atomic Research Center for translating it into English.